\documentclass[aps,pre,onecolumn,nofootinbib,notitlepage]{revtex4-1}
\usepackage{amsmath,amssymb,amstext,graphicx,color,xfrac,braket,accents,tabulary,booktabs,xcolor}
\begin{document}
\baselineskip=14pt
\hfill CALT-TH-2017-045
\hfill

\vspace{0.5cm}
\thispagestyle{empty}

\title{Quantum Decimation in Hilbert Space: \\ Coarse-Graining without Structure}
\author{Ashmeet Singh}
\email{ashmeet@caltech.edu}
\author{Sean M. Carroll}
\email{seancarroll@gmail.com}
\affiliation{Walter Burke Institute for Theoretical Physics, California Institute of Technology, Pasadena, CA 91125}

\newcommand{\be}{\begin{equation}}
\newcommand{\ee}{\end{equation}}
\newcommand{\bea}{\begin{eqnarray}}
\newcommand{\eea}{\end{eqnarray}}
\newcommand{\hs}{\mathcal{H}} 
\newcommand{\hst}{\widetilde{\mathcal{H}}} 
\newcommand{\A}{\left[\mathcal{A}\right]}
\newcommand{\D}{\left[\mathcal{D}\right]}
\newcommand{\B}{\left[{\hat{\mathcal{W}}}\right]}
\newcommand{\W}{\left[\mathcal{W}\right]}
\newcommand{\G}{\left[G\right]}
\newcommand{\T}{\left[\mathcal{T}_{d}\right]}
\newcommand{\iso}{\dot{=}}
\newcommand{\tdec}{{\{\theta\}}}
\newcommand{\pdec}{{\{\phi\}}}
\newcommand{\spsi}{\ket{\psi}}
\newcommand{\psimu}{\ket{\psi^{(\mu)}}}
\newcommand{\Cmu}{\left[ C^{(\mu)} \right]}
\newcommand{\Y}{\left[ C \right]}
\newcommand{\Ymean}{\left[ \bar{C} \right]}
\newcommand{\deltaY}{\left[ \Delta C \right]}
\newcommand{\OD}{\left[ O_{D} \right]}
\newcommand{\phiM}{\left[ \Phi \right]}
\newcommand{\Dim}{\textrm{dim\,}}
\newcommand{\Tr}{\textrm{Tr\,}}

\newcommand{\draftnote}[1]{\textbf{\color{red}[#1]}}
\newcommand{\ashmeet}[1]{\textbf{\color{blue}{#1}}}

\begin{abstract}
We present a technique to coarse-grain quantum states in a finite-dimensional Hilbert space.
Our method is distinguished from other approaches by not relying on structures such as a preferred factorization of Hilbert space or a preferred set of operators (local or otherwise) in an associated algebra.
Rather, we use the data corresponding to a given set of states, either specified independently or constructed from a single state evolving in time.
Our technique is based on principle component analysis (PCA), and the resulting coarse-grained quantum states live in a lower dimensional Hilbert space whose basis is defined using the underlying (isometric embedding) transformation of the set of fine-grained states we wish to coarse-grain. 
Physically, the transformation can be interpreted to be an ``entanglement coarse-graining" scheme that retains most of the global, useful entanglement structure of each state, while needing fewer degrees of freedom for its reconstruction. 
This scheme could be useful for efficiently describing collections of states whose number is much smaller than the dimension of Hilbert space, or a single state evolving over time.
\end{abstract}

\maketitle

\section{Introduction}

One of the challenges of doing practical calculations in quantum mechanics is that Hilbert space is very big: the number of dimensions is exponential in the number of degrees of freedom. Furthermore, not all degrees of freedom are created equal; some might be microscopic or high-energy and hard to access, while others may be irrelevant to certain dynamical questions.
It is therefore very often useful to coarse-grain, modeling quantum systems defined by states on some Hilbert space $\hs$ by states in some lower-dimensional Hilbert space $\widetilde\hs$, under conditions where the coarse-grained dynamics suffices to capture important properties of the system.

In practice, coarse-graining procedures typically rely on the existence of structure in Hilbert space that exists as part of the quantum system under consideration, and uses that structure to define a renormalization group (RG) flow \cite{kadanoff1967,fisher1998,wilson1975,fisher1974,cardy1996,maris+kadanoff1978}.
For example, there might be a notion of emergent space \cite{cao_etal2017,Cotler:2017abq} and associated locality. We imagine some decomposition of Hilbert space into local factors,
\be
  \hs = \bigotimes_i\hs_i ,
  \label{decomp}
\ee
where the factors $\hs_i$ come equipped with some nearest-neighbor structure (specifying when $\hs_i$ and $\hs_j$ are nearby), typically characterized by the form of interactions between factors in the Hamiltonian. Then it makes sense to coarse-grain spatially, grouping together nearby factors, as in the classic block-spin approach to the Ising model \cite{kadanoff1967,migdal1975}. Alternatively, one might appeal to the energy spectrum of the Hamiltonian, constructing an effective theory of low-energy states by integrating out high-energy ones. 

In quantum information theory, data compression has received a lot of attention over the last few decades and a considerable amount of work has been done. A lot of motivation for such techniques comes from quantum computation, and many different approaches have been suggested, including but not limited to Schumacher's data compression \cite{schumacher}, one-shot compression techniques \cite{oneshot}, the Johnson-Lindenstrauss lemma \cite{johnson_lindenstrauss,RSA:RSA10073} and its limitations in quantum dimensional reduction \cite{harrow_JL}, by application of elementary quantum gates \cite{plesch_etal}, and even considering overlapping qubits \cite{2017arXiv170101062C}, amongst others \cite{rozema_etal,qdatecompression_vaccaro_etal} (and references therein).

In this paper we pursue a different road to coarse-graining.
We imagine that we are given some particular set of states (or one state as a function of time) in Hilbert space, but no preferred notion of locality or energy, or a preferred factorization into individual degrees of freedom. Our specific motivation comes from quantum gravity and quantum cosmology, where notions of locality and energy are more subtle than in traditional laboratory settings, but the technique might be of wider applicability. Our method represents another technique in the  literature on compression quantum information and coarse-graining, but with an emphasis on the fact that the construction is based solely on structure of a set of given quantum states, without relying on any additional, preferred structure in Hilbert space.
In particular, we use principle component analysis (PCA) \cite{pcaref} to use a set of states $\{|\psi^{(\mu)}\rangle\} \in \hs$ to define a vector subspace $\hst \subset \hs$, such that a coarse-graining map onto $\widetilde\hs$ captures the most important information about the original states (in a sense we make precise below). We refer to this procedure as ``quantum decimation in Hilbert space'': a scheme where we coarse-grain quantum states by decimating or discarding irrelevant features determined by the structure of the states themselves, without presuming additional structure on Hilbert space. (Our method is distinct from past usage of the term ``quantum decimation'' in the literature \cite{chen+tuthill1985,castellani1981,matsubara+totsuji+thompson1991}, which refers to application of RG ideas to spin chains, Hubbard models and the like.) The idea of the PCA is to express the information contained in the original states in the most efficient way possible, by identifying the basis vectors along which most of the variation occurs,
and attaching a systematic notion of the relative importance of different basis vectors. This helps us identify global, important features of the state (determined by the states themselves) and physically one can relate this to preserving most of the relevant entanglement structure of the states (in any arbitrarily associated tensor factorization to Hilbert space). 

The paper is organized as follows. In Section (\ref{sec:PCAconstruct}), we construct the principle component basis for the set of fine-grained states we wish to coarse-grain, and define a PCA compression map which removes redundancy in the basis used to describe our states. In Section (\ref{cg_decimation}), we develop the details of the coarse-graining isometry based on decimation of the PCA expansion and discuss the physical question of ``what are we coarse-graining?'' and a possible application of the procedure in coarse-graining time evolution of a system. In section (\ref{conclusion}), we compare with other conventional coarse-graining schemes and data compression techniques in quantum information, and conclude.

\section{Constructing the Principle Component Basis}
\label{sec:PCAconstruct}

\subsection{Setting the Stage}

Consider a finite-dimensional Hilbert space $\hs$ of dimension $D = \Dim\hs$, equipped with a global basis $\{\ket{i} \}$ with $i = 1,2,\cdots, D$. 
``Global'' indicates that the basis spans all of $\hs$, and this choice of this basis is left arbitrary at this stage. Typically, this global basis can be identified with a tensor product structure which identifies degrees of freedom corresponding to subsystems. While in any practical setup such as many-body theory or quantum computation, one typically assumes a highly non-generic and preferred tensor factorization of $\hs$ based on the Hamitonian \cite{Tegmark:2014kka, carroll+singh_toappear} where features like locality and classical emergence might be manifest, we do not assume any such preferred structure. Our technique, at the ``data-compression" stage, works even without associating a tensor product structure to Hilbert space, but its interpretation (which we offer in section \ref{subsec:decimation_entanglement}) relies on the existence of an arbitrary factorization $\hs = \bigotimes_{j} \hs_{j}$, not necessarily corresponding to a quasi-classical one. A normalized state $\spsi \in \hs$ can be expanded in this global basis,
\begin{equation}
\label{psi_general}
\ket{\psi} \: = \: \sum_{i=1}^{D} c_{i} \ket{i} \: ,
\end{equation}
with $c_{i} = \braket{i | \psi}$ and $\sum_{i=1}^{D} |c_{i}|^2 = 1$.

Now imagine that we are given $M$ states in $\hs$, labeled by $\{\psimu\}$, $\mu = 1,2,\cdots,M$, which we will call the \emph{specifying states}.
Our goal is to harness the structure of these specifying states in $\hs$ to construct a coarse-graining procedure that will allow us to project them down to a subspace $\widetilde\hs$ that preserves as much relevant information as possible.
Each $\psimu$ be expanded in the chosen global basis,
\begin{equation}
\label{psimu_expansion}
\psimu \: = \: \sum_{i=1}^{D} c^{(\mu)}_{i} \ket{i} \: ,
\end{equation}
with $\sum_{i=1}^{D} |c^{(\mu)}_{i}|^2 = 1$. 
It will be convenient to package these components as a $(D \times 1)$ column vector, which we call $\Cmu$, 
\begin{equation}
\label{psimu_matrix}
 \left[ C^{(\mu)} \right]_{D \times 1}= \begin{bmatrix}
       c^{(\mu)}_{1}           \\[0.4em]
       c^{(\mu)}_{2}   \\[0.4em]
       .		            \\[0.1em]
       . 				\\[0.1em]
       . 				\\[0.1em]
       c^{(\mu)}_{D}
     \end{bmatrix} \: .
\end{equation}
We now bundle together the coefficients of all of the specifying states into a matrix $\Y$, of order $(D \times M)$, which we call our \textit{augmented matrix}:
\begin{equation}
\label{Ymatrix}
\Y_{D \times M} \: = \: \begin{bmatrix}
       C^{(1)}; & C^{(2)}; & \cdots ; & C^{(M)}           \\
     \end{bmatrix}
     \equiv
     \begin{bmatrix}
       c^{(1)}_{1}  & c^{(2)}_{1}  & \cdots  & c^{(M)}_{1}          \\[0.3em]
        c^{(1)}_{2}  & c^{(2)}_{2}  & \cdots   & c^{(M)}_{2}          \\[0.3em]
       . & .  & \cdots   & .          \\[0.3em]
              .  & .  & \cdots   & .          \\[0.3em]
        c^{(1)}_{D}  & c^{(2)}_{D}  & \cdots   & c^{(M)}_{D}          \\[0.3em]
     \end{bmatrix} \: .
\end{equation}

\begin{table}[t]\caption{List of Important Notation Used} 
\begin{center}
     \begin{tabular}{r  c | p{14cm} } \hline
     
\toprule
$\hs$ & \: \: \:  & Hilbert space of (fine-grained) dimension $D = \Dim\hs$. \\
$\{ \ket{i} \}$ & \: \: \:  & Global basis of $\hs$, $i = 1,2,\cdots,D$. \\
$\{ \psimu \}$ & \: \: \:  & Set of $M$ specifying states in $\hs$, $\mu = 1,2,\cdots, M$. \\
$\hst_{(d)}$ & \: \: \:  & Hilbert space of (coarse-grained) dimension $d = \Dim\hst_{(d)}$ with $\hst_{(d)} \subset \hs$ and $d \leq (M+1) < D$. \\
$\Cmu$ & \: \: \:  & Column-vector containing coefficients $c^{(\mu)}_{i}$ of $\psimu$ in the global $\ket{i}$ basis.  \\
$\left[C \right]$ & \: \: \:  & Augmented Matrix containing all $M$ specifying states as column vectors.\\
$\bar{C}^{(\mu)}$ & \: \: \:  & Mean value of the coefficients of $\psimu$ in the global basis. \\
$\OD$ & \: \: \:  & Un-normalized, uniform superposition state in $\hs$.  \\
$\deltaY$ & \: \: \:  & Augmented Matrix of deviation of each state from its mean value.  \\
$\{ e_{k} \}$ & \: \: \:  & Set of non-zero singular values of $\deltaY^{\dag} \deltaY$ and $\deltaY \deltaY^{\dag}, \: \: k = 1,2,\cdots,M$.   \\
$\left[\hat{\Phi} \right]$ & \: \: \:  & Un-normalized PCA basis vectors organized as a matrix. \\
$\left[\hat{\mathcal{W}} \right]$ & \: \: \:  & Un-normalized PCA weights organized as a matrix. \\
$\left[{\Phi} \right]$ & \: \: \:  & Normalized PCA basis vectors organized as a matrix. \\
$\left[{\mathcal{W}} \right]$ & \: \: \:  & Normalized PCA weights organized as a matrix. \\
$\{ \ket{\phi_{j}} \}$ & \: \: \:  & PCA basis vectors in $\hst_{(d)}$ with $j = 0,1,\cdots,(d-1)$. \\
$ \hat{\Pi}_{(d)}$ & \: \: \:  & Projection Map from fine-grained states in $\hs$ to coarse-grained states in $\hst_{(d)}$. \\
$ \T $ & \: \: \:  & Truncation matrix of order $d \times (M+1)$ with $\T_{ab} = \delta_{ab}$ \\
$\left[G_{d}\right]$ & \: \: \:  & Net coarse-graining transformation defined as $\left[G_{d}\right] = \T \left[{\Phi} \right]^{\dag}$. \\
\hline
\bottomrule
    \end{tabular}
\end{center}
    \label{tab:TableOfNotation}
\end{table}

The basic idea of coarse-graining is to reduce the effective dimensionality of Hilbert space, thus giving an effective description of the state, while retaining the global or large-scale physics of the state. Our current structure does not assume any notion of space or any associated notion of locality, or indeed any specific Hamiltonian. All we are working in is Hilbert space and an associated global basis. An idea of coarse-graining in such a setup would need to be equipped with the understanding of ``What are we coarse-graining?" and ``What are we losing under such a transformation?" since our regular ideas of spatial scales, lattices and locality are not present in the current scheme. This allows us to construct a more general prescription using Hilbert space ideas, which does not assume any preferred decomposition into subsystems or preferred observables, local or otherwise. These ideas are further discussed in Sections (\ref{subsec:decimation_entanglement}) and (\ref{conclusion}). 

We propose to perform principle component analysis (PCA) on the specifying states as a technique to reduce the dimensionality of our Hilbert space, thus resulting in a coarse-grained description for the specifying states. The resulting PCA coarse-graining prescription will be useful to coarse-grain the same set of specifying states only (unless there is some relationship between a separate state and the specifying states).
The PCA transforms the input into a set of linearly uncorrelated principle components, thus reducing any redundancy in describing the specifying states.
As is common in any PCA application, the first step is to remove the column-wise mean of the matrix $\Y$, which helps to isolate the sources of variance in the set of specifying states.
A mean-subtracted input allows the PCA components to have variance in reconstruction over and above the mean in a systematic way, where the $k$th component is more important in adding back variance as compared to the $(k+1)$st component. It is worth pointing out at this stage and as we will see, in our use of the PCA, the mean subtraction will be an important step in our physical interpretation of the coarse-graining transformation. 

 Let us begin by subtracting off the column-wise mean from the structure of our specifying states $\{\psimu\}$ described by the augmented matrix $\Y$. Let ${\bar{C}^{(\mu)}}$ be the mean of the $(D \times 1)$ column vector $\Cmu$, which is also the $\mu$th column of $\Y$,
 \begin{equation}
 \label{cmu_mean}
 {\bar{C}}^{(\mu)} \: = \: \frac{1}{D} \sum_{j = 1}^{D} c^{(\mu)}_{j} \: .
 \end{equation}
 We also define an un-normalized, uniform superposition state whose representation in the global $\{\ket{i} \}$ basis is the $(D \times 1)$ column vector $\OD$ with all entries equal to unity,
  \begin{equation}
  \label{OD}
  \OD_{D \times 1} \: \equiv \: \begin{bmatrix}
       1           \\[0.4em]
       1   \\[0.4em]
       .		            \\[0.1em]
       . 				\\[0.1em]
       . 				\\[0.1em]
       1
     \end{bmatrix}   \: .
  \end{equation}
While this uniform state $\OD$ is basis-dependent, we will argue in Section (\ref{cg_isometry+expectation}) that the relative inner product structure between the specifying states will be invariant under the coarse-graining for \emph{any} choice of global basis. Each choice of basis lends its own features which will be taken into account by the coarse-graining prescription, while at the same time, keeping the relative structure of the states invariant and offering a uniform interpretation in terms of entanglement coarse-graining for any associated tensor product structure to the chosen basis.

  Based on this, one can define the mean augmented matrix $\Ymean$ as the following $(D \times M)$ matrix,
  \begin{equation}
  \label{ymean}
  \Ymean \: = \:  \begin{bmatrix}
       {\bar{C}^{1}} O_{D}; & {\bar{C}^{(2)}} O_{D}; & \cdots ; & {\bar{C}^{(M)}} O_{D}           \\
     \end{bmatrix} \: ,
  \end{equation}
  and thus, the $\mu$th column of $\Ymean$ is simply,
  \begin{equation}
  \label{Ymean_mu}
  {\bar{C}^{(\mu)}} O_{D} \: = \: \frac{1}{D} \sum_{j = 1}^{D} c^{(\mu)}_{j} \begin{bmatrix}
       1           \\[0.4em]
       1   \\[0.4em]
       .		            \\[0.1em]
       . 				\\[0.1em]
       . 				\\[0.1em]
       1
     \end{bmatrix}   \: .
  \end{equation}
  One can now define the deviation of each of the specifying states from their respective means as,
  \begin{equation}
  \label{deltaY}
  \deltaY_{D \times M} \: = \: \Y_{D \times M} \: - \: \Ymean_{D \times M} \: ,
  \end{equation}
which will serve as a description of our states $\{\psimu\}$ 
based on the deviations of the coefficients from the mean ${\bar{C}^{(\mu)}}$ of each of the specifying states.
  
  \subsection{Implementing the Principle Component Analysis}
  \label{pcaimplement}
  
Starting with $M$ specifying states $\{\psimu\}$ in the $D$-dimensional Hilbert space $\hs$, we have decomposed them into a set of mean values organized into a matrix $\Ymean_{D \times M}$ and a set of deviations $\deltaY_{D \times M}$. In what follows, we focus on the case with $D > M+1$ (the ``+1" to become clear later), \textit{i.e.} with fewer states than the dimension of the space they live in. This is usually the relevant case, since state vectors describing physical systems commonly live in very large Hilbert spaces and the number of states one might wish to understand is much smaller. In the other limit with more states than dimensions, one would generically need the full support of the Hilbert space to describe them and a PCA based coarse-graining technique may not be very useful. The matrix $\deltaY$ captures all the information there is in our set of specifying states in the choice of basis, modulo the mean of each state which just adds a uniform contribution along each of the basis directions. We can think of $\deltaY$ as characterizing the deviation of the state from being a uniform superposition (in the average sense), which as we will see, will be important in interpreting the technique as an entanglement coarse-graining under any associated tensor product structure $\hs = \bigotimes_{j}\hs_{j}.$

We now perform a principle component analysis on the matrix $\deltaY$, which is implemented via a singular value decomposition (SVD). While one can directly perform a PCA on the coefficient matrix $\Y$ and work out the technique on similar lines as described below, we feel that delineating these different contributions makes the process rather clear and better physically motivated. We decompose $\deltaY$ as,
\begin{equation}
\label{svd1}
\deltaY_{D \times M} \: = \: \A_{D \times D} \:  \D_{D \times M} \: \B_{M \times M} \: ,
\end{equation}
  where $\A$ and $\B$ are unitary matrices and $\D$ is a diagonal matrix with $M$ non-zero singular values $\{e_{k}, \: k = 1,2,\cdots, M \}$ of $\deltaY$ on the diagonal, 
\begin{equation}  
  \D_{D \times M} =      \begin{bmatrix}
       e_1  & 0  & \cdots  &0      \\[0.3em]
      0  & e_{2}  & \cdots   & 0         \\[0.3em]
       . & .  & \cdots   & .          \\[0.3em]
              .  & .  & \cdots   & .          \\[0.3em]
        0  & 0  & \cdots   & e_{M}         \\[0.3em]
        0  & 0  & \cdots   & 0         \\[0.3em]
        . & .  & \cdots   & .          \\[0.3em]
        0 & 0  & \cdots   & 0          \\[0.3em]
     \end{bmatrix} \: .
\end{equation}  
  These non-zero singular values are the square roots of the non-zero eigenvalues of $\deltaY^{\dag} \deltaY$ and $\deltaY \deltaY^{\dag}$.  Following standard PCA procedure, we arrange the singular values on the diagonal in $\D$ in descending order, which helps capture the systematic addition of variance by the PCA,
  \begin{equation}
  e_{1} \geq e_{2} \geq \cdots \geq e_{M}.
  \end{equation}
  
It is most convenient to write the deviations from the mean as
\be
\deltaY_{D \times M} \: = \: \hat{\phiM}_{D \times M} \: \B_{M \times M} \: ,  
\ee
where the $D\times M$ matrix
\be
 \hat{\phiM}_{D \times M} \equiv \A_{D \times D} \:  \D_{D \times M}
\ee
defines the PCA basis, and the $M\times M$ matrix $\B$ defines the (un-normalized) PCA weights. The hat symbol $\left(\hat{.}\right)$ here is used to stress that the variable is not normalized. The use of hat $\left(\hat{.}\right)$ to denote operators, whenever used, will be clear from  context.
Written explicitly,
  \begin{equation}
  \label{pca1}
  \deltaY \: = \: \begin{bmatrix}
       \hat{\phi}_{1}; & \hat{\phi}_{1}; & \cdots ; & \hat{\phi}_{M}           \\
     \end{bmatrix}_{D \times M}      \begin{bmatrix}
       \hat{w}^{(1)}_{1}  & \hat{w}^{(2)}_{1}  & \cdots  & \hat{w}^{(M)}_{1}          \\[0.3em]
        \hat{w}^{(1)}_{2}  & \hat{w}^{(2)}_{2}  & \cdots   & \hat{w}^{(M)}_{2}          \\[0.3em]
       . & .  & \cdots   & .          \\[0.3em]
              .  & .  & \cdots   & .          \\[0.3em]
        \hat{w}^{(1)}_{D}  & \hat{w}^{(2)}_{D}  & \cdots   & \hat{w}^{(M)}_{D}          \\[0.3em]
     \end{bmatrix}_{M \times M} \: .
  \end{equation}
The columns $\left[ \hat{\phi}_{j} \right]_{D \times 1} \equiv e_{j} \left[ \mathcal{A}_{j} \right]_{D \times 1} \: \: j = 1,2,\cdots, M$ are the components in the original global $\{\ket{i} \}$ basis, of the $M$ new PCA basis vectors,
and $\hat{w}^{(\mu)}_{j}$ is the $j$-th un-normalized PCA weight for the specifying state $\psimu$. 

Thus, the deviation from the mean of $\psimu$ can be reconstructed as,
\begin{equation}
\label{deltaC}
\left[ \Delta C^{(\mu)} \right]_{D \times 1} \: = \: \sum_{j = 1}^{M} \hat{w}^{(\mu)}_{j} \left[ \hat{\phi}_{j} \right]_{D \times 1} \: .
\end{equation}
The columns of $\A_{D \times D}$, which we denote as $\left[ \mathcal{A}_{i} \right]_{D \times 1}, i = 1,2 \cdots,D$ form an orthonormal basis for the global Hilbert space $\hs$, since $\A$ is unitary, while just the first $M$ states in $\A_{D \times D}$ selected by the $M$ non-zero singular values $\{e_{k} \: , \: k = 1,2,\cdots, M\}$ are needed to form a complete basis for our specifying states $\psimu$ we wish to coarse-grain.
This step forms the information compression step: we have chosen a smaller set of linearly independent vectors who span a vector subspace that includes all of our specifying states $\psimu$. However, the scaling of each of these columns with the singular values $e_{j}$ to get $\left[ \hat{\phi}_{j} \right]$ renders the basis vectors un-normalized. Once this compression step is done, we can normalize our PCA basis states by associating the singular values with the PCA weights, by defining
\begin{equation}
\label{scaling1}
w^{(\mu)}_{j} = e_{j} \hat{w}^{(\mu)}_{j}, \: \: \: \:  j = 1, 2, \cdots, M \: \:  \:  \mathrm{and} \: \: \: \forall \: \mu \: .
\end{equation}
This lets us define the normalized PCA basis vectors as simply the first $M$ columns of the unitary $\A$,
\begin{equation}
\label{scaling2}
\left[ \phi_{j} \right]_{D \times 1} = \left[ \mathcal{A}_{j} \right]_{D \times 1}, \: \: \: \: j = 1, 2, \cdots, M \: \:  \:  \mathrm{and} \: \: \: \forall \: \mu \: .
\end{equation}
Thus, we have mapped the $D$ coefficients of each state $\Cmu$ to $M$ coefficients of the PCA expansion in the PCA basis $\phiM$ as obtained above, in addition to the mean coefficient of each state. To reconstruct the full state $\psimu$, we add the mean ${\bar{C}^{(\mu)}}$ multiplied by $\OD$ to obtain back $\Cmu$,
\begin{equation}
\label{recon1}
\psimu \: \equiv \Cmu =  \: {\bar{C}^{(\mu)}} \OD_{D \times 1} + \sum_{i = 1}^{M} w^{(\mu)}_{i} \left[ \phi_{i} \right]_{D \times 1} \: .
\end{equation}
In what follows, to avoid clutter in our equations, we drop the explicit use of the square brackets $ [ . ] $, which we have been using to denote matrices so far. 

The $\mu$th state $\psimu$ is normalized, hence we obtain the matrix representation of the normalization condition $\braket{\psi^{(\mu)} | \psi^{(\mu)}} = 1$ to be the following,
\begin{equation}
\label{norm1}
|{\bar{C}^{(\mu)}}|^{2} O^{\dag}_{D} O_{D} + \sum_{i = 1}^{M} \sum_{k = 1}^{M} w^{(\mu)}_{i} \left(w^{(\mu)}_{k}\right)^{*} \phi^{\dag}_{k} \phi_{i} + \sum_{k = 1}^{M} w^{(\mu)}_{k} O^{\dag}_{D} \phi_{k} + \sum_{k = 1}^{M} \left(w^{(\mu)}_{k}\right)^{*} \phi^{\dag}_{k} O_{D} \: = \: 1 \: .
\end{equation}
Before we further simplify the normalization condition, consider contracting the state $\psimu$ in Eq. (\ref{recon1}) with $\OD$,
\begin{equation}
\label{OD_phi1}
O^{\dag}_{D}  C^{{\mu}} \: = \: \sum_{j=1}^{D} c^{(\mu)}_{j} \: = \: {\bar{C}^{(\mu)}} O^{\dag}_{D} O_{D} + \sum_{k = 1}^{M}w^{(\mu)}_{k} O^{\dag}_{D} \phi_{k} \: .
\end{equation}
One can now use the fact that $O^{\dag}_{D} O_{D} = D$ and $\sum_{j=1}^{D} c^{(\mu)}_{j} = D {\bar{C}^{(\mu)}}$ to get,
\begin{equation}
\label{OD_phi2}
\sum_{k = 1}^{M} w^{(\mu)}_{k} O^{\dag}_{D} \phi_{k} \: = \: 0 \: = \: \sum_{k = 1}^{M} \left(w^{(\mu)}_{k}\right)^{*} \phi^{\dag}_{k} O_{D} .
\end{equation}
In addition to this, due to the mean subtraction in each column in Eq. (\ref{deltaY}), each of the PCA basis vectors $\left[ \phi_{j}\right], \: j = 1,2,\cdots, M$ has a zero mean $O^{\dag}_{D} \phi_{j} \: = \: \phi^{\dag}_{j} O_{D} \: = \: 0$. Hence, not only is the summand of Eq. (\ref{OD_phi2}) zero, but each term vanishes separately.
The PCA basis vectors $\left[\phi_{j} \right]_{D \times 1}$ are the columns of a unitary matrix, and are therefore orthonormal, $\phi^{\dag}_{k} \phi_{i} = \delta_{ik}$.
We can therefore use Eq. (\ref{OD_phi2}) to get the normalization condition for the $\mu$th state $\psimu$ as,
\begin{equation}
\label{norm}
|\sqrt{D} {\bar{C}^{(\mu)}}|^{2} + \sum_{k = 1}^{M} |w^{(\mu)}_{k}|^{2} \: = \: 1 \: \: \forall \: \mu \: = \: 1,2,\cdots,M \: .
\end{equation}
Thus, we have mapped the $D$ coefficients of each state $\psimu$ in the global basis to a mean value ${\bar{C}^{(\mu)}}$ and $M$ coefficients in the PCA basis $\phiM$, thus needing $M+1$ coefficients in this new basis to characterize the state. 

At this stage that we have captured the full information of each specifying state $\psimu$ in the $M+1$ coefficients and the constructed PCA basis. The dimensional reduction is not a result of integrating out small scale physics, rather it is simply a smart choice of basis, which minimizes redundancy in the description of our specifying states $\{\psimu\}$. We also know that $O^{\dag}_{D} \phi_{j} = 0$, making it orthogonal to all of the other PCA basis vectors, and is hence a linearly independent vector whose contribution is needed to reconstruct $\{\psimu\}$ from the $M+1$ coefficients. 
This motivates us to identify the ``zeroth'' component of the PCA basis $\phi_{0}$ and the corresponding PCA weight to be the mean contribution,
\begin{equation}
\phi_0 \equiv \frac{1}{\sqrt{D}}O_D\, , \quad w^{(\mu)}_{0} \equiv \sqrt{D}{\bar{C}^{(\mu)}} .
\end{equation}
The PCA basis now has $M+1$ basis and each contribution (mean, and otherwise) is treated homogeneously and one can express the basis set as $\left[ \Phi \right] = \left[ \phi_{0};\phi_{1};\cdots; \phi_{M} \right]$.
Thus we have (notice the sum runs from zero now),
\begin{equation}
\label{recon2}
\psimu  \: \equiv \: \Cmu \: = \:   \sum_{j = 0}^{M} w^{(\mu)}_{j} \left[ \phi_{j} \right]_{D \times 1} \: .
\end{equation}
Notice, we have added a factor of $\sqrt{D}$ to keep $\phi_{0}$ normalized like the other PCA basis vectors. Normalization of the state $\psimu$ is now simply written as, following Eq. (\ref{norm}),
\begin{equation}
\label{normnew}
\sum_{k = 0}^{M} |w^{(\mu)}_{k}|^{2} \: = \: 1 \: \: \forall \: \mu \: = \: 1,2,\cdots,M \: .
\end{equation}

\subsection{Mapping onto the PCA Subspace}
\label{subsection_PCAproj}

The PCA procedure discussed above provides us with $M+1$ vectors (the PCA basis) $\left[ \Phi \right]$, which span and act as a basis in a vector subspace containing our $M$ specifying states. Let us denote this subspace as $\hst_{(M+1)}$ with $\hst_{(M+1)}  \subset \hs$. For each of the PCA basis vector $\left[ \phi_{j}  \right], \: j = 0,1,\cdots,M$ we can identify the corresponding state vector $\ket{\phi_{j}} \in \hst_{(M+1)}$. This set of PCA vectors $\{ \ket{\phi_{j}} \}$ forms a complete, orthonormal basis set for $\hst_{(M+1)}$ and our specifying states can be expanded in this basis for $\hst_{(M+1)}$ following Eq. (\ref{recon2}),
\begin{equation}
\label{psimu_recon_M+1}
\psimu \: = \: \sum_{j = 0}^{M} w^{(\mu)}_{j}  \ket{\phi_{j}} \: .
\end{equation}
The \textit{j}-th basis state $\ket{\phi_{j}} \in \hst_{M+1}$ is embedded in the larger D-dimensional space $\hs$ and is connected to its representation in the global $\{ \ket{i}\} \in \hs$ basis via its matrix representation $\left[ \phi_{j} \right]$ of Eq. (\ref{scaling2}). 

Once this subspace $\hst_{(M+1)}$ has been defined and its basis identified, one can work with the specifying states exclusively in this subspace by mapping the state $\psimu$ from the larger space $\hs$ to $\hst_{(M+1)}$ by using an operator $\hat{\Pi}_{(M+1)}$. To understand the action of $\hat{\Pi}_{(M+1)}$ on the specifying states, we first connect the PCA weights $w^{(\mu)}_{j}$ with the global expansion coefficients $c^{(\mu)}_{i}$. For $\psimu$, by contracting both sides of $\sum_{i = 1}^{D}c^{(\mu)}_{i} \ket{i} = \sum_{j=0}^{M} w^{(\mu)}_{j}\ket{\phi_{j}}$ by $\bra{\phi_{k}}$   and using the orthonormality of the PCA basis $\braket{\phi_{k}|\phi_{j}} = \delta_{kj}$, we find
\begin{equation}
\label{w_spaceconnect}
w^{(\mu)}_{j} = \sum_{i=1}^{D} c^{(\mu)}_{i}\braket{\phi_{j}|i} \: .
\end{equation}
Thus, mapping to the $\hst_{(M+1)}$ space is achieved by,
\begin{equation}
\label{projM+1}
\hat{\Pi}_{(M+1)} = \sum_{j = 0}^{M}  \ket{\phi_{j}} \bra{\phi_{j}} \: .
\end{equation}
This of course keeps the specifying states unaltered, while mapping them onto the $\hst_{(M+1)}$ subspace with their expansion in the PCA basis, thus compressing the support needed to describe them. 
Also, any other vector $\ket{\alpha} \in \hst_{(M+1)} \subset \hs$ can be similarly mapped down from a D-dimensional to an $M+1$ dimensional space. While arbitrary states in $\hs$ not completely supported on $\hst_{(M+1)}$ can be mapped to $\hst_{(M+1)}$ using $\hat{\Pi}_{(M+1)}$, but such a map will non-systematically, and perhaps non-desirably, alter the structure of the state. 

Our focus in this paper is to coarse-grain the specifying states: the PCA map $\hat{\Pi}_{(M+1)}$ acts as the dimension compression step which can now be coarse-grained as described in Section (\ref{cg_decimation}).

\section{Coarse-Graining via Decimation}
\label{cg_decimation}
\subsection{Truncation of the PCA Expansion and Coarse-Graining}

 With this technology in hand, we can now explore how to systematically coarse-grain our states $\{\psimu\}$ to further lower-dimensional Hilbert spaces. 
With the PCA basis alone, we have already reduced the effective dimensionality of the underlying vector space from $D$ to $M+1$ using the PCA map $\hat{\Pi}_{(M+1)}$ without any loss in the description of the state, since the PCA simply chooses a smart basis which removes redundancy in their description. We now discuss the decimation prescription, in which we  truncate the PCA expansion of Eq. (\ref{psimu_recon_M+1}) as a method of coarse-graining, explicitly reducing the dimensionality of Hilbert space at the expense of throwing away certain features of the state.

Currently, a state $\psimu$ is expanded in the PCA basis $\phiM$, as done in Eq. (\ref{recon2}) {in the matrix representation describing its reconstruction in the global D-dimensional space $\hs$.} The $M$ non-zero singular values $e_{k} \: , \: k = 1,2,\cdots,M$ are arranged in descending order in the diagonal matrix $\D$ in Eq. (\ref{svd1}). {The PCA endows us with a systematic control of the contribution of different PCA components in reconstruction of the state.} Thus, the $j = 1$ component of the PCA, $w^{(\mu)}_{1} \phi_{1}$, carries \textit{maximum variance} in reconstructing the state $\Cmu$ over and above the zeroth component ($j=0$) \textit{i.e.} the state mean ${\bar{C}^{(\mu)}}$. The next $j = 2$ orthonormal component $w^{(\mu)}_{2} \phi_{2}$ has lesser variance than the $j=1$ component, and so on. The $k$-th component is more important that the $(k+1)$-st component in adding back variance over and above the mean to reconstruct the state. 

Since the tailing PCA components contribute little to the reconstruction of the state as compared to the preceding components, one could, depending on the required accuracy of reconstruction, neglect some of these tailing terms in the series to obtain an effective, coarse-grained description of the state. {To better understand relative importance of different PCA components in reconstructing the specifying states, one can look at the fractional contribution/importance ($\mathrm{Imp}$) of the $k$-th PCA component,}
\begin{equation}
\label{PCAimp}
\mathrm{Imp} (\phi_{k}) \: = \: \frac{e_{k}}{\sum_{j = 1}^{M} e_{j}} \: , \quad k = 1,2,\cdots,M \,.
\end{equation} 
Thus, in addition to the mean term $w^{(\mu)}_{0} \phi_{0} \equiv {\bar{C}^{(\mu)}} O_{D}$, one could choose the next $(d - 1)$ PCA terms with $1 \leq (d - 1) \leq M$ in the expansion as a coarse-grained of the state,
\begin{equation}
\label{truncate1}
\psimu_{CG(d)} \: \equiv \: \Cmu_{CG(d)} \: = \: \sum_{k = 0}^{d-1}w^{(\mu)}_{k} \left[\phi_{k}\right] \: ,
\end{equation}
where the contributions of the $k = d$ to $M$ components have been truncated and neglected. {In the above equation and in what follows, ``$CG(d)$'' indicates that the state has been coarse-grained (CG) to a $d$-dimensional reconstruction. The choice of $d$ can be made depending on the various fractional contributions (Eq. \ref{PCAimp}) of the PCA basis and the required accuracy of the coarse-grained description.} We have thus effectively mapped the $D$ coefficients of the state $\psimu$ in the original (global) basis to $d \leq (M+1) < D$ components in the truncated/coarse-grained PCA basis. 

Following the discussion in subsection (\ref{subsection_PCAproj}), we now construct a $d$-dimensional vector subspace $\hst_{(d)}$ with $\hst_{(d)} \subset \hst_{(M+1)} \subset \hs$, which covers the support of the $CG(d)$ coarse-grained specifying states. The first $d$ PCA vectors $\left[ \phi_{j} \right], \: j = 0,1,2,\cdots,d-1$ form an orthonormal basis for $\hst_{(d)}$ and can be identified with their corresponding set of basis state vectors $\ket{\phi_{j}}, \: j = 0,1,2,\cdots,d-1$. Before we construct the coarse-graining map, it is important to notice that truncating the PCA series renders the states un-normalized. Since we would like our coarse-grained vectors to be good quantum states satisfying probability summing to unity, we normalize the states by hand. A coarse-graining map $\hat{\Pi}_{(d)}$ can be constructed which projects and coarse-grains the state to $\hst_{(d)}$ and normalizes it as well,
\begin{eqnarray}
&\hat{\Pi}_{(d)} \: : \: \hs \to \hst_{(d)} \\
&\psimu \longmapsto \psimu_{CG(d)} = \frac{\sum_{j = 0}^{d-1}w^{(\mu)}_{j} \ket{\phi_{j}}}{||\sum_{k = 0}^{d-1}w^{(\mu)}_{k} \ket{\phi_{k}}||}.
\end{eqnarray}
As before, the basis states $\ket{\phi_{j}}$ are embedded in the original space $\hs$ via their matrix representations (Eq. (\ref{scaling2})). We see $2 \leq d \leq (M+1)$, and $d = 2$ is the most coarse-grained description of the specifying states as effective qubits, whereas the other limit $d = (M+1)$ takes it back to the full non-coarse-grained, albeit PCA-compressed description, as discussed in subsection \ref{subsection_PCAproj}. 
One can also define a series of nested subspaces 
\begin{equation}
\hst_{(2)} \subset \hst_{(3)} \subset \cdots \subset \hst_{(M+1)} \subset \hs,
\end{equation}
and a corresponding sequence of maps $\hat{\Pi}_{(d)}, \: d = 2,3,\cdots,M+1$, which progressively coarse-grain from just the PCA compression $(d = M+1)$ to a maximally coarse-grained description as an effective qubit $(d=2)$.

One can also consider a coarse-graining application where we admit non-normalized coarse-grained states, possibly due to inaccuracies in experimental setups or numerical precision. In that case, we can choose the coarse-grained dimension $d$ such that,
\begin{equation}
\label{CGnorm}
\sum_{k = 0}^{d-1} |w^{(\mu)}_{k}|^{2} \: = \: 1 - \epsilon ,
\end{equation}
for some $\epsilon$ small enough to not be detected experimentally or within numerical errors.


\subsection{The Coarse-Graining Isometry and Expectation Values}
\label{cg_isometry+expectation}

Let us recap what be have accomplished so far. We have coarse-grained each of our specifying states from a $D$-dimensional description in $\hs$ to a state living in the $d$-dimensional Hilbert space $\hst_{(d)} $with $d \leq M+1 <D$, and identified the $d$ expansion coefficients in the (truncated) PCA $\{ \ket{\phi_{i}} \} \: , \: i \: = \: 0,2,\cdots,(d-1)$ basis in $\hst_{(d)}$. Each of these basis states is connected to the fine-grained $D$-dimensional embedding in $\hs$ via its matrix representation, as found in Section (\ref{pcaimplement}). In this section, we aim to package our results and  formally define a transformation that directly relates the $d$ coarse-grained coefficients to the $D$ fine-grained coefficients. 

The PCA compression of $\psimu$ lives in $\hst_{(M+1)}$ and is described by $M+1$ coefficients $\{ w^{(\mu)}_{j}, \: j = 0,1,\cdots,M\}$. Let us denote this column by $\left[W^{(\mu)}\right]$, which is connected to the fine-grained description of the state $\psimu$ via the PCA basis $\phiM_{D \times (M+1)}$ following inversion of Eq. (\ref{recon2}) as,
\begin{equation}
\label{transform1}
\left[W^{(\mu)}\right]_{(M+1) \times 1} \: = \: \phiM^{\dag}_{(M+1) \times D} \Cmu \: .
\end{equation}
The PCA basis matrix, whose columns form an orthonormal basis in $\hst_{(M+1)}$, defines an isometric embedding, $\phiM^{\dag} \phiM = \mathbb{I}_{(M+1)}$, but in general $\phiM \phiM^{\dag} \neq \mathbb{I}_{D}$ as expected, where $\mathbb{I}_{p}$ is the $p$-dimensional identity. However, $\phiM \phiM^{\dag}$ acts as the identity in the subspace where our specifying states reside. This is tantamount to saying that the PCA projection $\hat{\Pi}_{(M+1)}$ leaves the specifying states invariant,
\begin{equation}
\label{identitysubspace}
\phiM \phiM^{\dag} \Cmu = \Cmu \: , \forall \mu = 1,2,\cdots,M \: .
\end{equation} 

Before describing truncation of the PCA series as an effective coarse-grained description of the state, it is instructive to understand how inner products of states are related in the two descriptions. Combining Eqs. (\ref{transform1}) and (\ref{identitysubspace}), it is easily seen that the inner product $\braket{\psi^{(\nu)} | \psi^{(\mu)}}$ is preserved while transforming from the global $D$-dimensional to the PCA $(M+1)$-dimensional description,
\begin{equation}
\label{innerprod_preserve}
\braket{\psi^{(\nu)} | \psi^{(\mu)}} = \left[C^{(\nu)} \right]^{\dag} \Cmu = \left[W^{(\nu)}\right]^{\dag} \left[W^{(\mu)}\right] \: .
\end{equation}
At this stage, one might worry about the basis-dependence of the PCA prescription outlined in Section (\ref{sec:PCAconstruct}), since the uniform, un-normalized state $\OD$ is a basis-dependent construction. Under different choices of global basis $\{ \ket{i} \}$ that lead to different augmented matrices $\Y$, one would end up with a different set of PCA basis vectors and weights, with the zeroth vector always identified as the uniform superposition state. However, this is not an issue since the relative inner product structure of the specifying states is invariant under change of global basis by a unitary transformation. This can be easily verified by using Eqs. (\ref{transform1}) and (\ref{identitysubspace}) for two different choices of global basis where the coefficients of the specifying states are connected by some unitary transformation $\left[ \Lambda \right]$. The PCA compression, while preserving overlaps between our set of specifying states in any arbitrary choice of basis, then also preserves the pairwise distances between states,
\begin{equation}
|| \ket{\psi^{\mu} - \psi^{\nu}} ||^{2} = \braket{\psi^{\mu} - \psi^{\nu}|\psi^{\mu} - \psi^{\nu}} = 2 - 2\: \mathrm{Re} \left( \braket{\psi^{\mu}|\psi^{\nu}} \right) \: ,
\end{equation}
and under truncation of the PCA expansion of Eq. (\ref{truncate1}), we preserve these overlaps and pairwise distances upto some error scale determined by the choice $d$ of the coarse-grained subspace.

The next step of coarse-graining via truncating the PCA expansion to the first $d$ coefficients of $\left[W^{(\mu)}\right]$ as a coarse-grained description of the state $\psimu$ can be achieved by a truncation matrix $\T$ that is of order $d \times (M+1)$ and is a diagonal matrix with ones on the diagonal, $\T_{ab} \: = \: \delta_{ab}$. Using this truncation matrix, the $d$ coefficients of the \emph{un-normalized} coarse-grained state $\psimu_{CG(d)}$, which we call $\left[W^{(\mu)}_{CG(d)}\right]$, can be obtained as,
\begin{equation}
\label{transform2}
\left[W^{(\mu)}_{CG(d)}\right]_{d \times 1} \: = \: \T \left[W^{(\mu)}\right] \: = \: \T \phiM^{\dag} \Cmu \: \equiv \left[G_{d}\right] \Cmu \: ,
\end{equation}
where we have defined the net coarse-graining transformation as $\left[G_{d}\right] = \T \phiM^{\dag}$, which satisfies $\left[G_{d}\right] \left[G_{d}\right]^{\dag} = \mathcal{I}_{d}$. This transformation captures both the PCA basis change and the truncation to retain the first $d$ components. Normalization of the coarse-grained state can be done by hand, as described in subsection (\ref{cg_decimation}).

 Another aspect is the behavior of expectation values of hermitian operators under our coarse-graining transformation. Consider a hermitian operator $\hat{\mathcal{O}} \in \mathcal{L}(\hs)$, which in the global basis for $\hs$ has a matrix representation $\left[ \mathcal{O} \right]_{D \times D}$, whose expectation in the $\mu$th state is
 \begin{equation}
 \label{Oexp_FG}
 \left< \hat{\mathcal{O}} \right>^{(\mu)}_{FG} = \left< \psi^{(\mu)} | \hat{\mathcal{O}} | \psi^{(\mu)} \right> \equiv \Cmu^{\dag} \left[ \mathcal{O} \right] \Cmu \: ,
 \end{equation}
 where the subscript $FG$ is to emphasize that we compute this expectation in the fine-grained, global description in $\hs$.
 One can construct the coarse-grained matrix representation of $\hat{\mathcal{O}}$ using our coarse-graining transformation $\left[G \right]$ as follows,
 \begin{equation}
 \label{O_CG}
 \left[ \mathcal{O}_{CG(d)} \right]_{d \times d} = \left[G_{d} \right] \left[ \mathcal{O} \right] \left[G_{d} \right]^{\dag} \: ,
 \end{equation}
 whose expectation value is computed with respect to the coarse-grained state $\left[W^{(\mu)}_{CG(d)}\right]$,
 \begin{equation}
 \label{Oexp_CG}
 \left< \hat{\mathcal{O}} \right>^{(\mu)}_{CG(d)} = \left[W^{(\mu)}_{CG(d)}\right]^{\dag}  \left[ \mathcal{O}_{CG(d)} \right] \left[W^{(\mu)}_{CG(d)}\right] = \Cmu^{\dag} \left(  \left[G_{d} \right]^{\dag} \left[G_{d} \right] \left[ \mathcal{O} \right] \left[G_{d} \right]^{\dag} \left[G_{d} \right] \right) \Cmu \: .
 \end{equation}
 Depending on how well/worse we decide to coarse-grain by choosing $d$, and the details and correlations in the specifying states, the coarse-grain expectations will differ from the fine-grained value, though the coarse-grained expectation $\left< \hat{\mathcal{O}} \right>^{(\mu)}_{CG(d)}$ approaches the fine-grained value as $d \to M+1$, and they are equal when $d = M + 1$.

\subsection{Decimation and Entanglement}
\label{subsec:decimation_entanglement}

Having developed a coarse-graining prescription based on a PCA transformation and further truncation of the expansion, our next task is to better understand what microscopic information is lost in the course of this transformation.
Ours is an unconventional coarse-graining prescription, since it is solely founded on the details of the quantum state given in some global basis. Most coarse-graining schemes assume more structure than this, be it a preferred split of the Hilbert space into tensor factors, a notion of locality in space, or energy modes beyond a certain cutoff that are to be integrated out. All we have is Hilbert space, a notion of a basis and a set of quantum states. A brief comparison of our PCA prescription with other conventional schemes will be done in Section (\ref{conclusion}). 

The basic question we wish to answer in this section is, what are we really losing when we perform the PCA and truncate the state description to retain the first $d$ components? What information are we discarding with the remaining $(M + 1 - d)$ components? 

To understand this, let us refer to the tensor product structure associated with the global fine-grained Hilbert space $\hs = \bigotimes_{j}\hs_{j}$. In most physical applications, one has a notion of subsystems, and correspondingly the global Hilbert space $\hs$ can be factorized preferentially as a tensor product of Hilbert spaces of each such subsystem. In what follows, we minimally assume some arbitrary tensor factorization of $\hs$, not necessarily equipped with some preferred decomposition governed by the Hamiltonian \cite{carroll+singh_toappear, Tegmark:2014kka, Piazza:2005wm} that might have notions of emergent space, locality, classical equations of motion, and the like. Our interpretation of the technique as an entanglement coarse-grainin just uses the existence of such a tensor product structure, and not it being special in any  particular way, though since we are working on more general grounds, our method can be adapted to more physically familiar cases.

For concreteness, let us associate a tensor product structure with $\hs$ of  $D = 2^n$ such that it can be thought of as the Hilbert space of $n$ qubits $\hs = (\hs_{2})^{\otimes n}$, where $\hs_{2}$ is the Hilbert space of a qubit. The argument which follows does not hinge on such a qubit factorization, but will work for any arbitrary factorization chosen. Let us write down the reconstruction of the $\mu$-th specifying state by explicitly writing out the mean term, the next $(d-1)$ terms being retained, and the $M + 1 - d$ truncated terms,
\begin{equation}
\label{cgphysics1}
\Cmu \: = \: {\bar{C}^{(\mu)}} \OD + \sum_{k = 1}^{d - 1}w^{(\mu)}_{k} \left[ \phi_{k} \right] + \sum_{l = d}^{M+1} w^{(\mu)}_{l} \left[ \phi_{l} \right] \: .
\end{equation}
 The mean term ${\bar{C}^{(\mu)}} \OD$ has by construction all the same entries. A state of $n$ qubits $\sim \OD$ represents a completely separable (product) state of the qubits, and thus has no entanglement between the constituent sub-factors. Thus, the mean state or the $\phi_{0}$ contribution sets a baseline state with the property of having no entanglement amongst its components. One can think of a different tensor structure to $\hs$ in terms of qudits, but the mean ${\bar{C}^{(\mu)}} \OD$ term still represents an unentangled state of the constituent sub-factors.
 
 The next $(d-1)$ terms in the PCA expansion of Eq. (\ref{cgphysics1}), $w^{(\mu)}_{k} \left[\phi_{k} \right] \: , \: k = 1,2,\cdots,(d-1)$ add most of the variance over and above the mean in reconstructing the (resultant, un-normalized) state. Thus, this sum of $(d - 1)$ terms adds most of the relevant entanglement structure of the state in the chosen tensor factorization of $\hs$. Of course, one may choose a factorization of $\hs$ under which the $\mu$-th specifying state may be unentangled to begin with and this argument of adding back relevant entanglement would not not be particularly useful. But for a generic decomposition, this understanding of entanglement coarse-graining would be a good notion of what our prescription is coarse-graining. The higher-order terms for $w^{(\mu)}_{l} \left[\phi_{l}\right] \: , \: l = d,d+1,\cdots,(M+1)$ have a  negligible (up to the coarse-graining scale set by choice of $d$) contribution in adding back variance to reconstruct the state, and hence also add minimal entanglement to the structure of the state in the chosen Hilbert space factorization. 

  \begin{figure}
  \label{pca_fig}
 \center
 \includegraphics[scale=0.60]{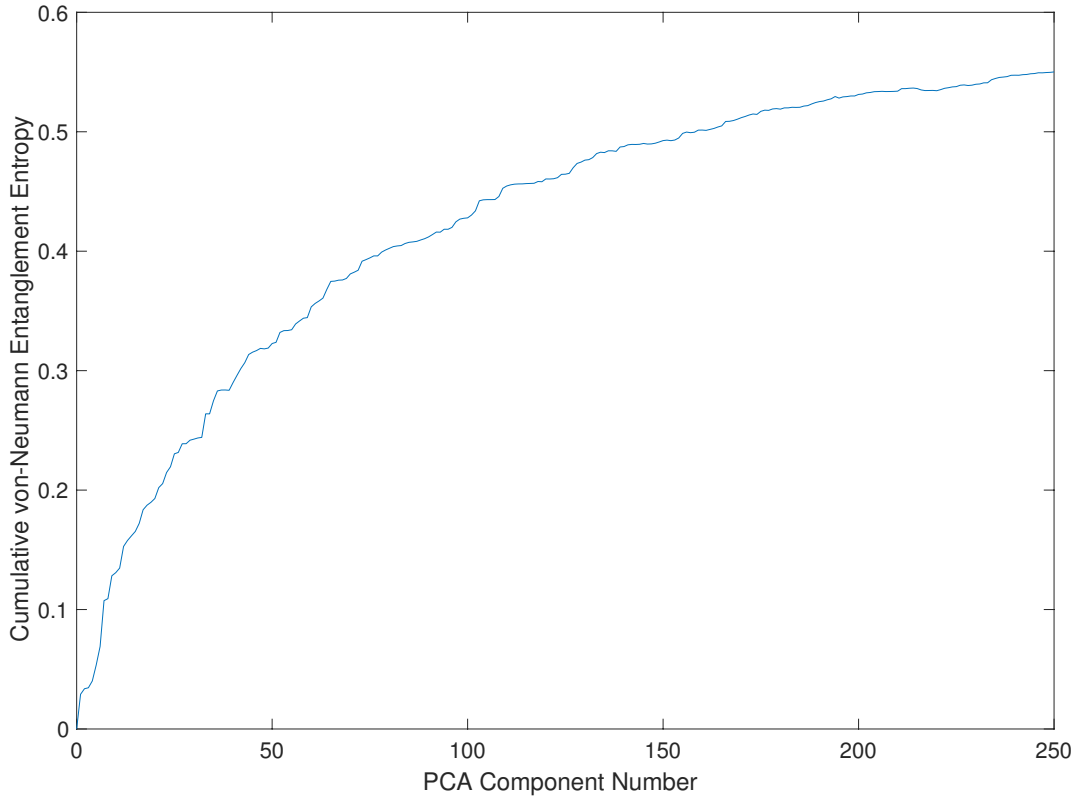}
 \caption{Plot of von~Neumann entanglement entropy of a constituent qubit of a state as a function of the number of PCA components retained in reconstructing the state.}
 \end{figure}
 
 As an example, we numerically constructed $M = 250$ specifying states of dimension $D = 2^{10}$. Each coefficient of these states was chosen from a pseudo-random distribution and then normalized. Following this, we performed our PCA-decimation procedure and reduced the dimensionality of each state to $d$ under the map $\hat{\Pi}_{(d)}$ (hence our coarse-grained states are normalized). The coarse-graining dimension $d$ was varied from $d = 1$, corresponding to retaining only the separable $\OD$ term, to $d = M+1$, corresponding to no truncation, only PCA compression. Now one can think of each state to be composed out of $n = 10$ qubits, and we can quantify the entanglement structure by looking at the von~Neumann entanglement entropy for these qubits in each of the specifying states. For instance, in the $\mu$-th specifying state $\psimu$, one can compute the entanglement entropy of the $q$-th qubit, $q = 1,2,\cdots, n$ (number of qubits) as
 \begin{equation}
 S^{(\mu)}_{q} = - \mathrm{Tr}_{q} \left( \hat{\rho}^{(\mu)}_{q} \mathrm{log}\,\hat{\rho}^{(\mu)}_{q} \right) ,
 \end{equation}
 where 
 \begin{equation}
 \hat{\rho}^{(\mu)}_{q} = \mathrm{Tr}_{\bar{q}} \left( \ket{\psi^{(\mu)}} \bra{\psi^{(\mu)}} \right) \: .
 \end{equation}
 
 Figure (1) plots the cumulative von~Neumann entanglement entropy of a chosen, constituent qubit in one of the constructed $M = 250$ states (again, chosen as a representative) as a function of the number of PCA components $d$ retained in reconstructing the state.  The idea of the plot is the saturation of the added entanglement entropy as one goes to more number of PCA components used to reconstruct the state. It is seen that only a few components are required to capture most of the entanglement structure while the higher orders have smaller contribution. By retaining the first $d$ components in the expansion, we recover the state $\psimu$  within error bounds (determined by the choice of $d$) in a way that we preserve the global or most relevant entanglement structure of the constituent sub-factors and lose irrelevant entanglement by truncating off the higher $ > d$ components. 
 
 The amount of correlations between the specifying states will be an important factor in determining how quickly the entanglement curve (as in Fig. 1) saturates. In general, for higher correlations amongst the specifying states, fewer PCA components would be required for most of the reconstruction, and one will expect a quick saturation in the entanglement build-up. In this sense, our PCA decimation coarse-graining prescription is related to entanglement coarse-graining, and is in the spirit of ignoring microscopic degrees of freedom and retaining large scale/global features; in this case, throwing away small, irrelevant entanglement while holding on the basic large scale structure of the state.
 
 \subsection{Coarse-Grained Time Evolution of a Quantum System}
\label{extensions}

We have based our coarse-graining prescription on very little structure in Hilbert space: equipped only with a global basis and a set of specifying states, our PCA decimation procedure maps states from a $D$-dimensional Hilbert space $\hs$ to $d<D$ dimensional space $\hst_{(d)}$ while retaining most of the global, relevant entanglement structure in the state (in some associated factorization). It is natural to ask in what setups can one adapt and put this coarse-graining prescription to use.

One possible application involves coarse-graining the discretized time evolution of a given initial state $\ket{\psi(t = 0)} \equiv \ket{\psi(0)} \in \hs$ of dimension $D = \Dim\hs$ with a global basis $\ket{i}, \: i = 1,2,\cdots,D $. The dynamics of states in $\hs$ are governed by some known and specified Hamiltonian $\hat{H}$. Consider unitarily evolving the initial state at $(M-1)$ time steps governed by some specified/chosen time step $\Delta t$, such that the state at the $j$-th time step $t_{j} = j \Delta t, \: j = 0,1,\cdots (M-1)$, is given by (we take units in which $\hslash = 1$),
\begin{equation}
\label{psi_t}
\ket{\psi(t_{j})} \: \equiv \:  \hat{U}(t_j) \ket{\psi(0)} \: = \: \exp{\left( - i\hat{H} t_{j} \right)} \ket{\psi(0)}, \: \: j = 0,1,2,\cdots,(M-1) \: .
\end{equation}
We thus have a collection of $M$ states living in a $D$-dimensional Hilbert space $\hs$. In the case when the number of time evolved states satisfy $D > (M+1)$, these $M$ states can act as our set of specifying states to undergo the PCA decimation prescription to be coarse-grained to a lower $d \left(\leq (M+1) < D\right)$ dimensional Hilbert space,
\begin{equation}
\label{timeevolve_state}
\ket{\psi(t_{j})}_{CG(d)} \: = \: \hat{\Pi}_{(d)}\ket{\psi(t_{j})}, \: \: j = 0,1,2,\cdots,(M-1) \: .
\end{equation}
If the Hamiltonian has desirable physical features such as locality, and if the time step is not too large, one would expect a high amount of correlation amongst the time evolved states. In this case, only a very few number of PCA basis components would be required to reconstruct the state. One can also find a coarse-grained representation of the Hamiltonian in the lower-dimensional $\hst_{(d)}$ space,
\begin{equation}
\label{HamCG}
 \left[ {H}_{CG(d)} \right]_{d \times d} = \left[G_{d} \right] \left[ {H} \right] \left[G_{d} \right]^{\dag} \: ,
\end{equation}
where $\left[ {H} \right]$ is the matrix representation of $\hat{H}$ in the global basis in $\hs$. Thus, using our PCA decimation prescription, one can compute a coarse-grained version of the time evolution of the state and use it as a proxy to study time-dependent features of the quantum system under consideration.

\section{Epilogue and Conclusion}
\label{conclusion}

Coarse-graining is a very important theme in understanding the behavior of realistic quantum systems which live in large Hilbert spaces of very large dimension. Many quantum coarse coarse-graining schemes \cite{white1992dmrg,vidal2007,matteo+soto+leuchs+grassl2017,busch+quadt1993,quadt+busch1994,faist2016,teo+rehacek+hradil2013,agon+balasubramanian+kasko+lawrence2014}  integrate out or eliminate irrelevant degrees of freedom to produce a coarse-grained description of the system. Renormalization Group techniques \cite{kadanoff1967,fisher1998,wilson1975,fisher1974,cardy1996,maris+kadanoff1978}  have been the cornerstone of coarse-graining ideas, and have proven to be extremely powerful and useful tools in physics. In particular, popular quantum coarse-graining schemes include Density Matrix Renormalization Group (DMRG) \cite{white1992dmrg,schollwock2005} and Entanglement Renormalization \cite{vidal2007} and their numerical implementations \cite{white1993,white+scalapino1998a,white+scalapino1998b,vidal2008,vidal2010,gaite2001,gaite2003,latorre+rico+vidal2004,osborne+nielson2002}. These, and many other coarse-graining schemes, assume substantial structure on Hilbert space. For instance, techniques like DMRG define an RG flow on the space of density matrices and serve as an effective truncation of Hilbert space of strongly correlated quantum many-body systems. Focusing on the low-energy properties of a system with a known Hamiltonian, one assumes a notion of spatial locality and  factorizability into state spaces on the lattice and numerical implementations further assume a preferred split into a system and an environment over which the trace is carried out to compute the properties at the level of the system. Similarly, in Entanglement Renormalization and its numerical implementations like MERA \cite{vidal2007}, one has a local lattice structure and aims to compute ground state properties for the system by defining a real space RG to dispose off small-distance degrees of freedom and entanglement (by the use of disentangling isometries, followed by block-decimation prescriptions). All coarse-graining schemes come equipped with an understanding of what global properties of the system one aims to retain, such as optimizing observable expectation values or correlation functions or entanglement between sub-systems; and which features are discarded which usually correspond to small scale entanglement, or high-energy modes etc. 

Techniques in quantum information theory to compress data and allow for dimensional reduction also form an interesting set of ideas to coarse-grain quantum information. Such schemes depend on the context at hand: for example, focusing on a typical subspace and ignoring its orthogonal complement, without much loss of fidelity, such as in Schumacher's noiseless quantum coding theorem, or compressing quantum information in a collection of qubits using elementary quantum circuit operations. Each technique has a specific aim and contextual validity, like the Johnson-Lindenstrauss lemma allows us to preserve pairwise distances up to a certain specified error tolerance and the dimension of the reduced subspace is then determined by this specified error and the number of points in the data set, and not on the dimensionality of the original space. Constructive implementations of the Johnson-Lindenstrauss lemma can be done via random projection and heavily relies on the Euclidean norm to measure pairwise distances, while on the other hand, dimensional reduction using PCA relies on specification of the dimension of the reduced subspace and projects onto a linear subspace. Thus, each technique has its range of validity and can be used depending on the physical system at hand.

While such methods are very useful, it is interesting to ask how one might coarse-grain a set of given quantum states in a Hilbert space which may or may not be associated with a Hamiltonian or the usual assumed structure on the space. In an effort in this direction, motivated by questions in quantum spacetime and emergent classicality, we have developed a coarse-graining prescription which uses Principle Component Analysis to first compress the dimensionality of Hilbert space by identifying a non-redundant basis (the PCA basis), followed by truncation of the last few PCA terms which contribute very little in reconstruction of the state. Physically, one can interpret this scheme as an entanglement coarse-graining (in some arbitrary associated factorization to Hilbert space) where, upon discarding the low importance terms, one only looses little and irrelevant entanglement structure of the state, while retaining major features in the reconstruction. One expects similarities between our PCA decimation scheme and other conventional coarse-graining prescriptions in the addition of appropriate structure. We feel this prescription is of a general nature, developed on a Hilbert space with very little structure, and can serve as a reliable means of first-principles quantum coarse-graining.

\section*{Acknowledgments}

We would like to thank Ning Bao, ChunJun (Charles) Cao, and Jess Riedel for helpful discussions during the course of this project. We are also thankful to an anonymous reviewer for their comments to help improve the manuscript. This material is based upon work supported by the U.S. Department of Energy, Office of Science, Office of High Energy Physics, under Award Number DE-SC0011632, as well as by the Walter Burke Institute for Theoretical Physics at Caltech and the Foundational Questions Institute. 

\bibliographystyle{utphys}
\bibliography{decimation-revised}

\end{document}